\documentclass{elsart3}
\usepackage{amsmath}
\usepackage{amssymb}
\usepackage{graphicx}% Include figure files
\usepackage{dcolumn}% Align table columns on decimal point
\usepackage{bm}% bold math
\begin{document}
\begin{frontmatter}
\journal{EXCON '04}

\title{Instability toward biexciton crystallization \\in one-dimensional electron-hole systems}%

\author{Kenichi Asano\corauthref{cor1}}
\ead{asano@phys.sci.osaka-u.ac.jp} \corauth[cor1]{Corresponding Author.
Tel.: +81-6-6850-5734, Fax: +81-6-6850-5351.}
\author{Tetsuo Ogawa}
\address{
Department of Physics, Osaka University, and CREST, JST,\\
1-1 Machikaneyama, Toyonaka, Osaka 560--0043, JAPAN
}


\begin{abstract}
One-dimensional (1D) electron-hole (e-h) systems in a high-density regime
is investigated by means of bozonization techniques.
It turned out that the systems are insulating even at the high density limit
and that the exciton Mott transition (insulator-to-metal transition) never occurs at absolute zero
 temperature.
The insulating ground state exhibits a strong instability
towards the crystallization of biexcitons.
\end{abstract}

\end{frontmatter}

\section{Introduction}
For more than three decades, electron-hole (e-h) systems realized in
strongly photo-excited semiconductors have been intensively studied both
theoretically and experimentally.
One of keywords to understand these systems is the exciton Mott
transition (insulator-to-metal transition), which has been believed
to take place with increasing of photo-excitation (e-h density).
The main origin of this transition is enhancement of the screening
effects which weaken the binding energy of the e-h bound
states \cite{Haug84}.
Such exciton Mott transition is confirmed both theoretically
and experimentally only in three-dimensional (3D) systems.
\par
Quite recently, extremely clean one-dimensional (1D) e-h systems are
realized experimentally in quantum wires fabricated in the T-shaped
quantum wells \cite{Yoshita04}.
From the theoretical point of view, such 1D systems are highly special
since at 1D an electron and a hole form an exciton bound state even
when the attractive e-h interaction is infinitesimally weak.
Therefore, we naively expect that the exciton Mott transition is absent
in 1D.
This argument above, however, is insufficient since the dynamical screening
effects are not considered.
\par
Some theoretical proposals on the topic of the exciton Mott transition in
1D has been reported in Ref.\cite{Wang01},
where a Bethe-Salpeter equation is solved approximately.
They found a critical e-h density, $n_{\rm c}$, where the exciton bound
state is dissociated, at absolute zero temperature even in 1D.
Nevertheless, the calculated optical absorption spectra show no evidence
of the exciton Mott transition, i.e., the exciton peak survives even for
densities higher than $n_{\rm c}$.
They claimed that this nonvanishing exciton peak explains some
experiments \cite{Ambigapathy97}.
Their results are, however, unreliable because their
perturbation theory based on the Fermi liquid picture, which is not
applicable to the 1D systems.
\par
With increasing the e-h density, the system goes to the weak-coupling
regime, where the characteristic interaction energy is much smaller than
the Fermi energies of electrons and holes.
In this regime, the bosonization method has advantages
in that it can take full account of the interaction processes.
The first application of this method to the 1D e-h systems
is given in Ref.\cite{Nagaosa93}, in which electrons
and holes are assumed to interact with each other through short-range force.
In the following, we apply this technique to the 1D e-h system with
long-range Coulomb interaction.
\par
Now, we turn to the properties of the insulating phase.
In 1D systems, the possibility of exciton crystallization is
theoretically pointed out \cite{Ivanov93} in the spinless e-h systems at
intermediate e-h densities, which is not expected to take place in 3D systems.
Here, ``crystallization'' means that the excitons are located at regular
intervals.
In the following, we will discuss the possibility of the biexciton
crystallization in 1D systems consisting of electrons and holes with
spin at absolute zero temperature.
\par
\section{Formulation}
In the following, we consider properties of the ``ground state'' of a one-dimensional e-h system.
Such consideration makes sense in comparison with the actual
experimental situation because the interband relaxation time is usually
long enough to realize intraband thermal quasiequilibrium.
\par
In the weak-coupling regime, the low-energy physics is unaffected by the
linearization of band dispersion, $\epsilon^{(\mu)}(k)\!=\!rv_{\rm
F}^{(\mu)}(k-rk_{\rm F})$, near the right ($r\!=\!+$) and the left ($-$)
Fermi points $k=rk_{\rm F}$, respectively, where $v_{\rm F}^{(\mu)}$
denotes the Fermi velocity and $\mu\!=\!{\rm e,h}$ is the index for the
electron and the hole, respectively.
When both electron and hole bands are far off the half filling, the
umklapp processes are negligible.
Then, the Hamiltonian can be written in the g-ology form, which is
characterized by the momentum-dependent coupling parameters
$g^{(\mu\mu')}_1(q)$, $g^{(\mu\mu')}_2(q)$ and $g^{(\mu\mu')}_4(q)$.
Here,  $g^{(\mu\mu')}_1$ specifies the backward scattering, while
$g^{(\mu\mu')}_2$ and $g^{(\mu\mu')}_4$ do the forward ones.
Note that there is interband (e-h) scattering process ($\mu \neq \mu'$)
as well as the intraband (e-e and h-h) scattering process ($\mu=\mu'$).
For simplicity, we restrict ourselves to the e-h symmetric case,
$v_{\rm F}\!=\!v^{\rm (e)}_{\rm F}\!=\!v^{\rm (h)}_{\rm F}$ and
$g_n^{\rm (ee)}\!=\!g_n^{\rm (hh)}$.
The conclusions obtained below are quite general and
qualitatively unchanged for the asymmetric cases.
More detailed and general calculations will be presented elsewhere
\cite{Asano}.
\par
In the quantum wire formed in T-shaped quantum wells, the interaction
potential behaves as unscreened Coulomb potential ($\propto 1/r$) at the
long interparticle distance $r$.
Thus, the coupling parameters of the forward scattering show a
logarithmic divergence at $q\rightarrow 0$ \cite{Schulz93}
In fact, they can be approximated in the form of $g_{2,4}^{\rm
(ee)}(q)\!=\!g_{2,4}^{\rm (hh)}(q)\!\sim\!g_{2,4}\!+\!2g_0K_0(|q|d)$ and
$g_{2,4}^{\rm (eh)}(q)\!=\!g_{2,4}^{\rm (he)}(q)\!\sim\!-g'_{2,4}-2g_0K_0(|q|d)$, using the modified Bessel function $K_0(x)$, and the effective
diameter of the quantum wire $d$.
The coupling constants $g_{2,4}$ and $g'_{2,4}$ specify the short-range
part of forward scattering, and $g_0\!=\!e^2/\epsilon\pi v_{\rm F}$
does its long-range part, where $e$ and $\epsilon$ denote the electron
charge and the dielectric constant, respectively.
On the other hand, the coupling parameters
of the backward scattering can be substituted to constants as
$g_1^{\rm (ee)}(q)\!=\!g_1^{\rm (hh)}(q)\!=\!g_1$ and
$g_1^{\rm (eh)}(q)\!=\!g_1^{\rm (he)}(q)\!=\!-g_1'$, because they show no
singularity at $q\!\sim\!\pm 2k_{\rm F}$.
\par
Now, we introduce the phase operators $\Phi_\nu^{(\xi)}(x)$ and
$\Theta_\nu^{(\xi)}(x)$ ($\xi\!=\!\pm,\ \nu\!=\!\rho,\sigma$) through
\begin{eqnarray*}
\partial_x\Phi^{(\pm)}_\rho\!&=&\!-\pi\sum_{r\sigma}\rho_{r\sigma}^{(\pm)},\;
\partial_x\Phi^{(\pm)}_\sigma\!=\!-\pi\sum_{r\sigma}(-1)^{\delta_{\sigma\downarrow}}
\rho_{r\sigma}^{(\pm)},\\
\partial_x\Theta^{(\pm)}_\rho\!&=&\!\pi\sum_{r\sigma}r\rho_{r\sigma}^{(\pm)},\;
\partial_x\Theta^{(\pm)}_\sigma\!=\!\pi\sum_{r\sigma}r(-1)^{\delta_{\sigma\downarrow}}
\rho_{r\sigma}^{(\pm)},
\end{eqnarray*}
where $\sigma\!=\!\uparrow,\downarrow$ denotes the spin,
$\delta_{\sigma\sigma'}$ is the Kronecker's delta,
$\rho_{r\sigma}^{(\pm)}\!=\!(\rho_{r\sigma}^{\rm
(e)}\pm\rho_{r\sigma}^{\rm (h)})/2$,
and $\rho^{(\mu)}_{r\sigma}(x)$ denotes the density at position $x$ for
the particle with indices $r$ and $\sigma$. It is noteworthy that
$\Phi^{(+)}_\rho$ and $\Phi^{(-)}_\rho$ are associated with the mass
and charge densities (the sum and difference of electron and hole
densities), respectively.
\par
Then, the g-ology Hamiltonian is written as
\begin{eqnarray}
&&{\mathcal H}=\sum_{\nu=\rho,\sigma,\xi=\pm}{\mathcal H}_\nu^{(\xi)}+{\mathcal H}_{\rm bs}^{\rm (intra)}+{\mathcal H}^{\rm (inter)}_{\rm bs},\nonumber\\
&&{\mathcal H}^{(\xi)}_\nu=\frac{v_\nu^{(\xi)}}{2\pi}\!\!\int\!\!dx
\Bigg\{\!K^{(\xi)}_\nu\!\!\left[\partial_x\Theta_\nu^{(\xi)}\right]^2
\!\!+\!\frac{1}{K^{(\xi)}_\nu}\!\left[\partial_x\Phi_\nu^{(\xi)}\right]^2\!\!\Bigg\}\nonumber\\
&&\mbox{\hspace{0.5truecm}}+\!\frac{v_{\rm F}}\pi\!\int\!\! dxdx'\frac{2g_0\delta_{\xi-}
\delta_{\nu\rho}}{\sqrt{(x-x')^2+d^2}}\partial_x\Phi^{(-)}_\rho\partial_{x'}\Phi^{(-)}_\rho,\nonumber\\
&&{\mathcal H}_{\rm bs}^{\rm (intra)}\!=\frac{v_{\rm F}g_1}{\pi\alpha^2}\!\!\int dx
 \cos 2\Phi_\sigma^{(+)}\cos 2\Phi_\sigma^{(-)},\nonumber\\
&&{\mathcal H}_{\rm bs}^{\rm (inter)}\!=\!-\frac{v_{\rm
 F}g_1'}{\pi\alpha^2}\!\!\!\int\! dx \sum_\xi\cos 2\Phi_\rho^{(-)}\cos 2\Phi_\sigma^{(\xi)}
\end{eqnarray}
Here, we used the cut-off constant $\alpha$, the renormalized velocity
$v_\nu^{(\xi)}\!=\!v_{\rm F}({a_\nu^{(\xi)}b_\nu^{(\xi)}})^{1/2}$ and coupling
constant $K_\nu^{(\xi)}\!=\!(a_\nu^{(\xi)}/b_\nu^{(\xi)})^{1/2}$,
where $a_\nu^{(\xi)}$ and $b_\nu^{(\xi)}$ are given as
$a^{(\pm)}_\rho\!=\!1+g_4-g_2+g_1/2\mp g_4'\pm g_2'$, $b^{(\pm)}_\rho\!=\!1+g_4+g_2+g_1/2\mp g_4'\mp g_2'$, $a_\sigma^{(\pm)}\!=\!1+g_1/2$, and $b_\sigma^{(\pm)}\!=1-g_1/2$.
\par
The field operators for the electron and the hole can also be expressed as
$\psi^{(\mu)}_{r\sigma}(x)\!=\!(2\pi\alpha)^{-1/2}\eta^{(\mu)}_{r\sigma}\exp[ir(k_{\rm F}x+\phi_{r\sigma}^{(\mu)})]$ in terms of the Klein factor
$\eta^{(\mu)}_{r\sigma}$ and the phase operators,
$\phi^{\rm (e)}_{r\sigma}\!=\!\sum_{\xi=\pm}(-\Phi_\rho^{(\xi)}\!-\!\sigma\Phi_\sigma^{(\xi)}\!+\!r\Theta_\rho^{(\xi)}\!+\!r\sigma\Theta^{(\xi)}_\sigma)/2$ and
$\phi^{\rm (h)}_{r\sigma}=\sum_{\xi=\pm}\xi(-\Phi_\rho^{(\xi)}\!-\!\sigma\Phi_\sigma^{(\xi)}\!+\!r\Theta_\rho^{(\xi)}\!+\!r\sigma\Theta^{(\xi)}_\sigma)/2$.
\par
\section{Results}
If we neglect the intra- and interband backward scattering terms,
${\mathcal H}_{\rm bs}^{(\rm intra)}$ and ${\mathcal H}_{\rm bs}^{(\rm
inter)}$, the Hamiltonian is decoupled into four parts,
each reduced to the Tomonaga-Luttinger (TL) form.
In this case, they can be exactly diagonalized and
the system is shown to have four gapless excitation modes corresponding
to four different degrees of freedom: $(\nu,\xi)\!=\!(\rho,+)$,
$(\rho,-)$, $(\sigma,+)$, and $(\sigma,-)$.
\par
We can also evaluate the asymptotic behavior of the various correlation
functions with the form, $C(x)\!=\!\langle {\mathcal
O}^\dagger(x){\mathcal O}(0)\rangle$.
After some calculations similar to those of Ref.\cite{Schulz93}, we find
that the correlation function of ${\mathcal O}_{\rm BED}\!=\!\psi^{\rm
(e)\dagger}_{+\uparrow}\psi^{\rm
(e)\dagger}_{+\downarrow}\psi^{\rm (h)\dagger}_{-\downarrow}\psi^{\rm
(h)\dagger}_{-\uparrow}\psi^{\rm (h)}_{+\uparrow}\psi^{\rm (h)}_{+\downarrow}\psi^{\rm
(e)}_{-\downarrow}\psi^{\rm (e)}_{-\uparrow}\!+\!{\rm h.c.}\!\sim\!\cos 4\Phi^{(-)}_\rho
$ show the slowest decay.
It is interesting to note that ${\mathcal O}_{\rm BED}$ appears in the slowly
varying ($0k_{\rm F}$) component of the biexciton density (BED) operator.
The biexciton creation operator here is given as
$\psi^{\rm (e)\dagger}_\uparrow\psi^{\rm (e)\dagger}_\downarrow\psi^{\rm
(h)\dagger}_\downarrow\psi^{\rm (h)\dagger}_\uparrow$
with $\psi^{(\mu)\dagger}_\sigma=\sum_r\psi_{r\sigma}^{(\mu)\dagger}$. 
This fact shows the strong tendency toward the biexciton formation. In
this sense, the ground state has the character of ``biexciton liquid.''
\par
Now, we turn to effects of the backward scattering terms.
It is noteworthy that intraband backward scattering is (marginally)
irrelevant if the interband one is absent.
Thus, we can neglect it to investigate the relevancy of the
interband backward scattering.
We treat the interband backward scattering term in the self-consistent
harmonic approximation (SCHA) and the renormalization group (RG) method.
The both results show that the interband backward scattering is always
relevant independent of $g_0\!>\!0$.
Because the coupling constant $g_1'$ is renormalized to a large value,
$\Phi_\rho^{(-)}$ and $\Phi_\sigma^{(\pm)}$ are fixed by the condition: $
\cos2\Phi_\rho^{(-)}\!=\!\cos2\Phi_\sigma^{(+)}\!=\!\cos2\Phi_\sigma^{(-)}\!=\!\pm
1$, which minimizes the interband backward scattering term.
\par
This localization of the phases leads to the energy gaps in the
excitation modes corresponding to $(\nu,\xi)\!=\!(\rho,-)$,
$(\sigma,+)$, and $(\sigma,-)$.
The point is that the charge excitation,
$(\nu,\xi)\!=\!(\rho,-)$, is gapful.
This results in that the system is an insulator even in the high e-h
density limit, and suggests that the exciton Mott transition is absent
at absolute zero temperature.
\par
There remains a single gapless excitation corresponding to
$(\nu,\xi)\!=\!(\rho,+)$, since ${\mathcal H}_\rho^{(+)}$
is decoupled from the other part of the Hamiltonian.
As will be mentioned below, its coupling constant $K_\rho^{(+)}$ specify
the power of the algeblaic decay of some correlation functions.
\par
Now, let us discuss the character of the insulating ground state
obtained above.
To this end, we investigate the asymptotic behavior of the
correlation functions $C(x)\!=\!\langle {\mathcal O}^\dagger(x){\mathcal O}(0)\rangle$.
There are two important points.
\begin{itemize}
\item[1.] Because the phases $\Phi_\rho^{(-)}$ and $\Phi_\sigma^{(\pm)}$
      are localized, the fluctuations of their conjugate operators
      $\Theta_\rho^{(-)}$ and $\Theta_\sigma^{(\pm)}$ diverge.
      Thus, $C(x)$ decays exponentially if ${\mathcal O}(x)$
      contains $\Theta_\rho^{(-)}$ or $\Theta_\sigma^{(\pm)}$.
\item[2.] If ${\mathcal O}(x)$ vanishes under the phase-fixing
	  condition, $C(x)$ shows an exponential decay.
\end{itemize}
Owing to these criteria, it is sufficient to consider the following two
operators, ${\mathcal O}_{\rm MDW}\!=\!\sum_{\mu
r\sigma}\psi^{(\mu)\dagger}_{r\sigma}\psi^{(\mu)}_{-r\sigma}\!\sim\!\cos(2k_{\rm
F}x\!-\!\Phi_\rho^{(+)})$ and ${\mathcal O}_{\rm
BEI}\!=\!\psi^{\rm (e)}_{r\uparrow}\psi^{\rm (e)}_{r\downarrow}\psi^{\rm
(h)}_{-r\downarrow}\psi^{\rm (h)}_{-r\uparrow}\!\sim\!\exp(2i\Theta_\rho^{(+)})$, 
where these expressions are derived using the condition
$\cos2\Phi_\rho^{(-)}\!=\!\cos2\Phi_\sigma^{(+)}\!=\!\cos2\Phi_\sigma^{(-)}\!=\!1$.
\par
The operator ${\mathcal O}_{\rm MDW}(x)$ denotes the $2k_{\rm F}$
oscillatory component of mass density wave (MDW), while ${\mathcal
O}_{\rm BEI}(x)$ annihilates a ``biexciton'' located at $x$.
The asymptotic forms of their correlation functions can be
explicitly evaluated as $C_{\rm MDW}(x)\!\sim\!x^{-{K_\rho^{(+)}}/2}$ 
and $C_{\rm BEI}(x)\!\sim\!x^{-2/{K_\rho^{(+)}}}$.
As a result, we can see that the formation of the $2k_{\rm F}$-MDW and
the Bose-Einstein condensation of biexcitons (biexcitonic insulator,
BEI), shows the strongest instability at $K_\rho^{(+)}\!<\!2$ and
$K_\rho^{(+)}\!>\!2$, respectively.
\par
In the weak coupling regime ($g_n\!\ll\!1$ and $g'_n\!\ll\!1$),
we obtain $K_\rho^{(+)}\!\sim\!1$, which leads to the strong instability
toward the formation of the $2k_{\rm F}$-MDW.
The formation of $2k_{\rm F}$-MDW can be interpreted as ``biexciton
crystallization.''
In fact, the $2k_{\rm F}$ charge density waves (CDWs) of the electron and
the hole are simultaneously formed in an in-phase way in the $2k_{\rm
F}$-MDW state since the fluctuation of the total charge density is
strongly suppressed.
As a result,  two electrons and two holes are effectively accumulated to
form ``biexcitons'' arranged regularly with the periodicity $\pi/k_{\rm F}$.
\par
\section{Summary and Discussion}
%When the interband backward scattering term, i.e. the origin of the
%charge gap, is neglected, the metallic ground state exhibits a
%``biexiton liquid'' character, due to the long-range nature of the
%interaction.
%This is quite contrast to the results for the e-h TL model with
%short-range interaction \cite{Nagaosa93}, in which the $2k_{\rm
%F}$-CDW of the hole is the most dominant order.
%\par
By means of the SCHA and RG techniques, we examined the relevancy of the
interband backward scattering, and showed that it is always relevant and
the charge gap $\Delta$ opens.
This means that the system is an insulator even in the high e-h density
regime in 1D, and indicates the absence of the exciton Mott
transition.
We also investigated the character of this insulator and found the
strong instability toward the biexciton crystallization.
\par
Our results clearly show the importance of biexciton correlation,
whereas the previous theories take into account only the exciton
correlations.
It was also shown that only the collective modes are significant, that
is the special feature of the 1D system.
This fact also indicates that the traditional criterion for the
exciton Mott transition is not applicable to the 1D e-h system, in which the
formation of the bound state between a single electron and a single hole is
considered using screened interaction \cite{Wang01}.
\par
In experiments, we can expect that the temperature $k_{\rm B}T$ is
larger than the energy gap $\Delta$.
In this case, the effects of the backward scattering terms are negligible.
Then, the system shows the character of ``biexciton liquid,'' within the
length scale $\sim v_{\rm F}/k_{\rm B}T$.
This result is quite contrast to that for the e-h TL model with short-range interaction \cite{Nagaosa93}, in which the $2k_{\rm F}$-CDW of the hole is the most dominant order.
The ``biexciton-liquid'' character is consistent with the recent
observation of the photoluminescence spectra \cite{Yoshita04};
they show a broad peak at high e-h density, which continuously connects
with the biexciton peak at low e-h density.

\end{document}